\documentclass[preprint,showpacs,preprintnumbers,amsmath,amssymb]{revtex4}

\usepackage{graphicx}
\usepackage{dcolumn}
\usepackage{bm}
\usepackage{epstopdf}

\newcount\driver
\newcount\bozza

scaled\magstep1                  
scaled\magstep1 \font\msytw=msbm10 scaled\magstep1
 
scaled\magstep1                 \font\indbf=cmbx10 scaled\magstep2

scaled \magstep2

{\count255=\time\divide\count255 by 60
\xdef\hourmin{\number\count255}
        \multiply\count255 by-60\advance\count255 by\time
   \xdef\hourmin{\hourmin:\ifnum\count255<10 0\fi\the\count255}}

\let\a=\alpha \let\b=\beta         \let\d=\delta     \let\e=\varepsilon
  \let\h=\eta      \let\k=\kappa     \let\l=\lambda
                          \let\r=\rho
\let\s=\sigma             
\let\ps=\psi   \let\o=\omega     
        \let\L=\Lambda    
             
\let\O=\Omega

\def\EE{{\cal E}}\def\VV{{\cal V}}
\def\WW{{\cal W}}

\def\LL{{\cal L}}
\def\DD{{\cal D}}

\def\pp{{\bf p}}\def\qq{{\bf q}}\def\xx{{\bf x}}
   
\def\yy{{\bf y}}\def\kk{{\bf k}}\def\nn{{\bf n}}
\def\zz{{\bf z}}

       \def\oo{{\underline \omega}}
\def\ee{{\underline \varepsilon}}

\def\NNN{\hbox{\msytw N}}

        \def\EE{\hbox{\msytw E}}

\let\==\equiv

\let\io=\infty
\let\0=\noindent

\def\*{{\hfill\break\null\hfill\break}}

\def\tilde#1{{\widetilde #1}}

\def\tende#1{\,\vtop{\ialign{##\crcr\rightarrowfill\crcr
             \noalign{\kern-1pt\nointerlineskip}
             \hskip3.pt${\scriptstyle #1}$\hskip3.pt\crcr}}\,}
\def\otto{\,{\kern-1.truept\leftarrow\kern-5.truept\to\kern-1.truept}\,}

\def\wh#1{\widehat{#1}}
\def\hat#1{\wh{#1}}
\def\sqt[#1]#2{\root #1\of {#2}}

\def\bp{{\bar \ps}}

\def\EE{{\cal E}}\def\VV{{\cal V}}
\def\WW{{\cal W}}

\def\LL{{\cal L}}
\def\DD{{\cal D}}

\def\T#1{{#1_{\kern-3pt\lower7pt\hbox{$\widetilde{}$}}\kern3pt}}
\def\VVV#1{{\underline #1}_{\kern-3pt
\lower7pt\hbox{$\widetilde{}$}}\kern3pt\,}
\def\W#1{#1_{\kern-3pt\lower7.5pt\hbox{$\widetilde{}$}}\kern2pt\,}
\def\Re{{\rm Re}\,}

\def\indica{\leaders \hbox to 0.5cm{\hss.\hss}\hfill}
\def\guida{\leaders\hbox to 1em{\hss.\hss}\hfill}
\mathchardef\oo= "0521

\def\pp{{\bf p}}\def\qq{{\bf q}}\def\xx{{\bf x}}
\def\yy{{\bf y}}\def\kk{{\bf k}}\def\nn{{\bf n}}
\def\zz{{\bf z}}

\def\oo{{\underline \omega}}

\def\qed{\raise1pt\hbox{\vrule height5pt width5pt depth0pt}}
 
  \def\bp{{\bar p}} 

\def\indic{\hbox{\raise-2pt \hbox{\indbf 1}}}

\def\NNN{\hbox{\msytw N}}

\def\ins#1#2#3{\vbox to0pt{\kern-#2 \hbox{\kern#1 #3}\vss}\nointerlineskip}

\newdimen\xshift \newdimen\xwidth \newdimen\yshift
\newcount\griglia

\def\insertplot#1#2#3#4#5#6{%
\xwidth=#1pt \xshift=\hsize \advance\xshift by-\xwidth \divide\xshift by 2%
\begin{figure}[ht]
\vspace{#2pt} \hspace{\xshift}
\begin{minipage}{#1pt}
#3 \ifnum\driver=1 \griglia=#6
\ifnum\griglia=1 \openout13=griglia.ps \write13{gsave .2
setlinewidth} \write13{0 10 #1 {dup 0 moveto #2 lineto } for}
\write13{0 10 #2 {dup 0 exch moveto #1 exch lineto } for}
\write13{stroke} \write13{.5 setlinewidth} \write13{0 50 #1 {dup 0
moveto #2 lineto } for} \write13{0 50 #2 {dup 0 exch moveto #1
exch lineto } for} \write13{stroke grestore} \closeout13
\includegraphics{griglia.ps} \fi
\includegraphics{#4.ps}\fi%
\ifnum\driver=2 \fi
\end{minipage}
\caption{#5}
\end{figure}
}

\newdimen\shift \shift=-1.5truecm
\def\lb#1{%
\ifnum\bozza=1
\label{#1}\rlap{\hbox{\hskip\shift$\scriptstyle#1$}}
\else\label{#1} \fi}

\def\be{\begin{equation}}
\def\ee{\end{equation}}
\def\bea{\begin{eqnarray}}\def\eea{\end{eqnarray}}
\def\bean{\begin{eqnarray*}}\def\eean{\end{eqnarray*}}
\def\bfr{\begin{flushright}}\def\efr{\end{flushright}}
\def\bc{\begin{center}}\def\ec{\end{center}}
\def\bal{\begin{align}}\def\eal{\end{align}}
\def\ba#1{\begin{array}{#1}} \def\ea{\end{array}}
\def\bd{\begin{description}}\def\ed{\end{description}}
\def\bv{\begin{verbatim}}\def\ev{\end{verbatim}}
\def\nn{\nonumber}
\def\Halmos{\hfill\vrule height10pt width4pt depth2pt \par\hbox to \hsize{}}
\def\pref#1{(\ref{#1})}

\def\ins#1#2#3{\vbox to0pt{\kern-#2 \hbox{\kern#1 #3}\vss}\nointerlineskip}

\newdimen\xshift \newdimen\xwidth \newdimen\yshift
\newcount\griglia

\def\insertplot#1#2#3#4#5#6{%
\xwidth=#1pt \xshift=\hsize \advance\xshift by-\xwidth \divide\xshift by 2%
\begin{figure}[ht]
\vspace{#2pt} \hspace{\xshift}
\begin{minipage}{#1pt}
#3 \ifnum\driver=1 \griglia=#6
\ifnum\griglia=1 \openout13=griglia.ps \write13{gsave .2
setlinewidth} \write13{0 10 #1 {dup 0 moveto #2 lineto } for}
\write13{0 10 #2 {dup 0 exch moveto #1 exch lineto } for}
\write13{stroke} \write13{.5 setlinewidth} \write13{0 50 #1 {dup 0
moveto #2 lineto } for} \write13{0 50 #2 {dup 0 exch moveto #1
exch lineto } for} \write13{stroke grestore} \closeout13
\includegraphics{griglia.ps} \fi
\includegraphics{#4.ps}\fi%
\ifnum\driver=2 \fi
\end{minipage}
\caption{#5}
\end{figure}
}

\newdimen\shift \shift=-1.5truecm
\def\lb#1{%
\label{#1}\rlap{\hbox{\hskip\shift$\scriptstyle#1$}}
\else\label{#1} \fi}

\def\be{\begin{equation}}
\def\ee{\end{equation}}
\def\bea{\begin{eqnarray}}\def\eea{\end{eqnarray}}
\def\bean{\begin{eqnarray*}}\def\eean{\end{eqnarray*}}
\def\bfr{\begin{flushright}}\def\efr{\end{flushright}}
\def\bc{\begin{center}}\def\ec{\end{center}}
\def\bal{\begin{align}}\def\eal{\end{align}}
\def\ba#1{\begin{array}{#1}} \def\ea{\end{array}}
\def\bd{\begin{description}}\def\ed{\end{description}}
\def\bv{\begin{verbatim}}\def\ev{\end{verbatim}}
\def\nn{\nonumber}
\def\Halmos{\hfill\vrule height10pt width4pt depth2pt \par\hbox to \hsize{}}
\def\pref#1{(\ref{#1})}

\usepackage{amsmath}
\usepackage{amsfonts}
\usepackage{amssymb}
\usepackage{slashbox}
\usepackage{epstopdf}

\font\msytw=msbm9 scaled\magstep1 
scaled\magstep1

\let\a=\alpha \let\b=\beta    \let\d=\delta
\let\e=\varepsilon
  \let\h=\eta    \let\k=\kappa \let\l=\lambda
                 \let\r=\rho
\let\s=\sigma     
\let\ps=\Psi   \let\o=\omega
   \let\L=\Lambda 
         
\let\O=\Omega 

\def\EE{{\cal E}} \def\VV{{\cal V}}
 \def\WW{{\cal W}}
 
\def\LL{{\cal L}}  
\def\DD{{\cal D}}

\def\qq{{\bf q}} \def\pp{{\bf p}}
 \def\xx{{\bf x}} \def\yy{{\bf y}} \def\zz{{\bf z}}
\def\kk{{\bf k}}

\def\nn{\nonumber}

\def\NNN{\hbox{\msytw N}}

\def\\{\hfill\break}
\def\={:=}
\let\io=\infty
\let\0=\noindent

\def\tende#1{\,\vtop{\ialign{##\crcr\rightarrowfill\crcr\noalign
{\kern-1pt
    \nointerlineskip} \hskip3.pt${\scriptstyle #1}$\hskip3.pt\crcr}}\,}
\def\otto{\,{\kern-1.truept\leftarrow\kern-5.truept\to\kern-1.truept}\,}

\def\wh{\widehat}
\def\to{\rightarrow}

\def\qed{\hfill\raise1pt\hbox{\vrule height5pt width5pt depth0pt}}

\def\be{\begin{equation}}
\def\ee{\end{equation}}
\def\bp{\begin{pmatrix}}
\def\ep{\end{pmatrix}}
\def\bea{\begin{eqnarray}}
\def\eea{\end{eqnarray}}
\def\nn{\nonumber}
\def\pref#1{(\ref{#1})}

\def\lb{\label}

\begin{document}

\title{Universality, exponents and anomaly cancellation in disordered Dirac fermions}
\author{Vieri Mastropietro}

\affiliation{Dipartimento di Matematica F. Enriques,
Universit\'a di Milano, Via C. Saldini 50, 20133, Milano, Italy
}


%

\begin{abstract}
Disordered 2D chiral fermions
provide an effective description of several materials including graphene and
topological insulators. While previous analysis considered
delta correlated
disorder and no ultraviolet cut-offs, we consider here the effect of short range correlated disorder
and the presence of a momentum cut-off, providing a more realistic description of condensed matter models.
We show that the density of states is anomalous with a critical exponent function
of the disorder and that 
conductivity is universal
only when the ultraviolet cut-off is removed, as consequence 
of the supersymmetric cancellation of the anomalies.
\end{abstract}

 \maketitle

\section{Introduction}

It is known that several materials
exhibit fermionic excitations with linear dispersion relation close to the Fermi level,
which can be effectively described in terms of $2+1$dimensional Dirac fermions.
Early examples include systems displaying integer quantum Hall effect
\cite{LFSG} and {\it d-wave superconductors} \cite{NTW,ASZ}
and more recently 
{\it graphene}
\cite{V3,H0,Mi,CG,Z1,FCO} and {\it topological insulators} \cite{N1,Mo}.
In particular, in the case of graphene at half filling it has been observed
\cite{N2} that the {\it optical} conductivity
(for frequencies greater than the temperature) 
is essentially
constant in a wide range of frequencies and very close to the {\it universal} value
$(\pi/2)(e^2/h)$, which also happens to be the value
found for the system of non-interacting $2d$ Dirac fermions \cite{LFSG}, a remarkable result in view of the fact that interactions are not particularly weak.
In transport measurements an {\it universal} value for the conductivity is also found, of order 
of the conductivity quantum 
$e^2/h$ \cite{N3}; again a surprising result in view of the presence of disorder which is surely relevant in such experiments.

It is of course important to understand if and under which conditions such universality 
can be understood theoretically. In presence of weak short range interactions, after first
perturbative computations claiming non vanishing corrections, it
was finally rigorously proved \cite{GMP} that the optical conductivity is {\it exactly equal} to its non interacting value. Note that the emerging description is in terms of a Nambu-Jona Lasinio model
and the natural cut-off provided by honeycomb lattice
ensures the correct symmetries and allows the proof 
of the complete cancellation of the interaction corrections.
On the other hand, in the case of {\it long range} Coulomb interaction
it has been predicted that the optical conductivity is still equal to the non interacting value
\cite{H0}, the argument this time being based on the divergence of the Fermi velocity.
However, the Fermi velocity divergence found in the Coulomb case at very low 
frequencies is clearly rather unphysical,
and simply signals ultimate inadequacy of the usual model of instantaneous  Coulomb interaction. With the increase of the Fermi velocity the
retardation effects eventually become important, so that the retarded current-current interaction must be added to the
Coulomb density-density interaction; the emerging model is in this case $QED_{4,3}$ (with an ultraviolet  cut-off ) in which the fermionic velocity is  different from the light velocity.
Such system have been analyzed before in
\cite{GGV} , \cite{GMPgauge} and it was found that
the flow of the Fermi velocity stops at the velocity of light $c$, and, maybe most importantly,
that the coupling constant (i. e. the charge) in the theory is {\it exactly} marginal (anomalous critical exponents are found); as a consequence of that, the optical conductivity
is {\it not} equal to its non interacting value but corrections are found \cite{HM}, which are however
quite small and still universal at lowest order (they depend only only from the fine structure constant). 

When we turn to the analysis of the effect of disorder on the conductivity, the natural emerging description is 
in terms of {\it disordered Dirac fermions}, which were extensively analyzed
along the years. In the case of {\it chiral preserving disorder} it was found that the density of states is vanishing with a critical exponent (non trivial function
of the disorder strength) but the conductivity is {\it universal} and not depending from the disorder amplitude,
see \cite{LFSG,NTW}. Such results, obtained using the replica trick, were confirmed and extended by a
Supersymmetric analysis of such models \cite{ASZ,Mu1,Mu2} leading to a functional integral in
Bosonic and Grassmann variables and a {\it local} quartic interaction.
It is  rather natural to relate such results to the universal conductivity found by transport measurements
in graphene
\cite{Mi}, despite the understanding of why the dominant disorder in graphene should 
preserve chirality is an open issue which may be related to how the sample is produced. 
However, even assuming that disorder preserves chirality, several questions still remain to be understood.
The results in \cite{LFSG, NTW,ASZ,Mu1,Mu2} on Dirac fermions with disorder
where found assuming {\it delta correlated}
disorder and an unbounded  fermionic dispersion relation (no ultraviolet cut-offs). Such features makes an exact analysis possible (even
non perturbative, see \cite{Mu2} and references therein)  but produce {\it ultraviolet divergences}
similar to the one present in local Quantum Field Theory in $d=1+1$ (for instance in the Thirring model),
which could lead to some discrepancy with respect to lattice 
models
(see \cite{Z0,Z} and the discussion in \cite{Zirn}) which are of course free from ultraviolet divergences.
As the dispersion relation  (in graphene or in the other condensed matter applications)
is approximately conical ("relativistic")
only in a small region around the Fermi level, it is natural to consider the presence of a  
momentum cut-off; moreover, 
a {\it short-range} correlated disorder is a much more realistic description
for condensed matter systems, see {\it e.g.} \cite{FCO,N1,Mo}. 
Both such features make disordered Dirac fermions {\it free}
from ultraviolet divergences,  and it is 
therefore natural to ask if the results 
with no cut-off and $\d$-correlated disorder are sufficiently robust to 
persists under the above more realistic conditions. Our main results are the following:
\begin{enumerate}
\item In the case of short range disorder, if the momentum cut-off is removed 
the density of states vanishes with a critical exponent and the conductivity is universal;
that is, the system has the same qualitative behavior than the case of $\d$-correlated disorder.
\item If the momentum cut-off is not removed, the density of states is still anomalous but the
conductivity has in general disorder-dependent corrections.
\end{enumerate}
Therefore, the vanishing of the density of states with an anomalous exponent is a robust property 
for chiral disordered fermions, but 
the {\it exact} vanishing of the disorder corrections to the conductivity does not survive in general to the presence of a momentum cut-off. From a Renormalization Group point of view this is rather natural. In presence
of chiral disorder the theory is {\it marginal} with a line of fixed points; therefore corrections are expected, as in the case of the optical conductivity in presence of e.m. interaction. From this perspective, it is the {\it absence} 
of corrections the more surprising feature of disordered Dirac fermions with no cut-off; as it will be clear
from the subsequent analysis, it is a direct consequence of the validity of the Adler-Bardeen theorem
and the exact cancellation of the chiral anomaly due to the supersymmetry, which is valid only
when the momentum cut-off is removed. The presence of corrections to the conductivity
in presence of an ultraviolet cut-off
 may have of course implications for the physics of graphene, in which a natural ultraviolet cut-off is provided by the honeycomb lattice.

The presence of momentum cut-off and of non local disorder prevents the use of any {\it exact}
methods, like the ones adopted in  \cite{LFSG, NTW,ASZ,Mu1,Mu2}, and  
one has therefore to rely
on functional integral methods, which are more lengthy but of more general applicability.
In particular we will use 
multiscale methods based on  Wilsonian Renormalization Group (RG), in the more advanced form
used in constructive Quantum Field Theory, see e.g. \cite{GJ}. 
Such form is exact, in the sense that the irrelevant terms (in the technical RG sense)
are fully taken into account, while 
in most non exact RG  implementations the irrelevant terms are simply neglected; as non local disorder
or finite cut-offs are irrelevant in the infrared regime, non exact RG cannot distinguish
between local and non local disorder, or the presence or absence of an ultraviolet cut-off. 

Using the 
supersymmetric formalism we can rewrite disordered Dirac fermions
in terms of functional integrals.
The fermionic sector strongly reminds the {\it non local Thirring model} which was constructed using a multiscale analysis respectively 
in \cite{Le, M2} for the ultraviolet problem 
and in \cite{M3} for the infrared part; therefore restricting to the fermionic sector 
a full non-perturbative construction of the model can be achieved, in the sense of a proof of the well definiteness of the functional integrals removing cut-offs; this would be parallel to \cite{DZ}, in which the restriction to the bosonic sector of an
hyperbolic sigma model coming by a disordered electron system was constructed. 

The plan of the paper is the following. In \S 2 we define the model and we explain its supersymmetric representation. In \S 3 we analyze the critical theory at $E=0$, we derive Ward Identities
and we show the validity of the Adler-Bardeen theorem and the supersymmetric cancellation of the anomalies in the limit of removed ultraviolet momentum cut-off;
also, the relation with universality will be explained. In \S 4 we consider the non critical theory $E\not=0$
and we discuss the infrared problem. Finally, in \S 5 the main conclusions are discussed.

\section{Disordered Chiral fermions  and Supersymmetric representation}

\subsection{The Dirac equation with vector disorder}

The (regularized) first quantized  Hamiltonian describing chiral disordered Dirac fermions is
\be
H=\sum_{i=1}^2  \s_i (i\tilde\partial_{i}+g A_i(\xx))\label{11}
\ee
with $\xx=(x_1,x_2)\in \L_a$, $\L_a$ is a square lattice with step $a$ with 
antiperiodic boundary conditions, $\s_i$ are Pauli matrices
$$
\s_1=\begin{pmatrix} 0  & 1 \\ 1\0 &  0 \end{pmatrix}\quad\quad \s_2=
\begin{pmatrix} 0  & -i\\ i\0 &  0 \end{pmatrix}\quad\quad \s_0=
\begin{pmatrix} 1  & 0\\ 0 &  -1 \end{pmatrix}
$$
and $\tilde\partial_i$ is a regularized (smeared) derivative
\be \tilde\partial_i f_\xx=\sum_\yy \chi^{-1}(\xx-\yy)\partial_i
f_{\yy} \quad\partial_{i}f(\xx)={1\over 2 a}(f(\xx+a{\bf
e}_i)-f(\xx-a {\bf e}_i))
\ee
with $\chi(\xx)$ is a cut-off function defined as the Fourier transform of $\hat\chi(\kk)$, with
$\hat\chi(\kk)$ a smooth function  which is $\hat\chi(\kk)=0$ for
$|\kk|\ge 2^{N+1}$
and $\hat\chi(\kk)=1$ for   $|\kk|\le 2^{N}$; $A_i(\xx)$ is a Gaussian random field with short range (but non local) correlation
\be
\mathbb{E}(A_{i}(\xx)A_j(\yy))=\d_{i,j}v(\xx-\yy)
\ee
and
\be
|v(\xx-\yy)|\le C e^{-\k|\xx-\yy|}\label{alal}
\ee
and we will set $\k=1$ for definiteness.

One is mainly interested in the average of the two-point function, from which the density of states can be computed
\be
\mathbb{E}[<\xx|{1\over i H- E}|0>]\label{s1}
\ee
and in the average of the product of two functions
\be
\mathbb{E}[<\xx|{1\over i H- E}|0>_{+-} <0|{1\over i H+ E}|\xx>_{+-}]\label{s2}
\ee
which is related to the conductivity.
In the absence of disorder
\be
 <\xx|{1\over i H- E}|0>
={1\over L^2}\sum_{\kk}e^{i\kk\xx}\hat\chi(\kk)\begin{pmatrix} -E   & {1\over a}[i\sin a k_1+\sin a k_2] \\ {1\over a}[i\sin a k_1-\sin a k_2]  &  -E
\end{pmatrix}^{-1}\label{pro}
\ee
In the following we will assume that $2^{N}<<{\pi\over a}$ in order to avoid the {\it fermion doubling}
problem. Indeed 
at  $E=0$ the denominator in the r.h.s. of \pref{s1}
is vanishing, in the $L\to\io$ limit, not only at $\kk=(0,0)$ but also
at $\kk=(0,\pi/a),(\pi/a,0),(\pi/a,\pi/a)$ modulo $2\pi/a$. The condition 
$2^{N}<<{\pi\over a}$ ensures that
the only remaining pole is the one at $\kk=0$, so  preventing the fermionic species multiplication
but at the same time preserving the chiral symmetry. The role of the lattice cut-off
is just to make the functional integrals appearing below  well defined and it will removed first.

%
%
%
%
%
%
%


\subsection{Supersymmetric formalism}

It is well known, see for instance \cite{Mu1},
that the average of the two-point function \pref{s1} and the average of the product
of two functions \pref{s2} can be represented in terms of a supersymmetric
functional integral in the chiral basis. 
One introduces a finite set of Grassmann
variables $\psi^+_{\o,\xx},\psi^-_{\o,\xx}$ with $\o=\pm$ and defines, if $\e=\pm$, the Grassmann integration by
\be
\int d\psi^\e_{\o,\xx} \psi^\e_{\o,\xx}=1\quad\quad\int d\psi^\e_{\o,\xx}=0
\ee
where  $d\psi^\e_{\o,\xx}$ is another set of Grassmann variables. Therefore if
$
\DD\psi=\prod_{\xx,\o=\pm}d\psi^+_{\o,\xx}d\psi^-_{\o,\xx}$
we can write
\be <\xx| [{1\over i H- E}]|\yy>|_{\o,-\o'}={\int \DD\psi e^{-(\psi^+,  A \psi^-) }
\psi^-_{\xx,\o}\psi^+_{\yy,\o'}\over \int \DD\psi e^{-(\psi^+, A \psi^-) }} \label{aaxx}\ee
where $A=\s_1( i H- E)$. 
Note that the denominator of \pref{aaxx} is equal to 
${\rm Det } A$, and that 
$\int \DD\phi  e^{-(\phi^+, A \phi^-)}={1\over {\rm Det }A}$,
for any $n\times n$ invertible matrix $A$ with $Re A>0$ and 
$\phi^+,\phi^-$ complex numbers with $(\phi^+)^*=\phi^-$.
Therefore one obtains the following representation of the average of the two point function
\be G_{\l,E,N;\o,\o'}(\xx)
=\mathbb{E}[ \int \DD\phi\DD\psi
[\psi^-_{\o,\xx}\psi^+_{\o',\yy}] e^{-(\psi^+, A \psi^-)-(\phi^+, A \phi^-)} ]\label{88}\ee
and integrating over the disorder, calling $\Psi^+=(\psi^+_+,\psi^+_-,\phi^+_+,\phi^+_-)$
and $\Psi^-=(\psi^-_+,\psi^-_-,\phi^-_+,\phi^-_-)$, $\phi^+=(\phi^-)^*$
\be
 G_{\l,E,N;\o,\o'}(\xx)=\int P(d\psi) P(d\phi) e^{\VV}
\psi^-_{\o,\xx}\psi^+_{\o',0}\label{66}
\ee
where, if $\l=2g^2$
\be
\VV=-
\l \int d\xx d\yy v(\xx-\yy)\sum_{\a,\a'=\phi,\psi}\sum_{\o=\pm}\Psi^+_{\a,\o,\xx}\Psi^-_{\a,\o,\xx}
\Psi^+_{\a',-\o,\yy}\Psi^-_{\a',-\o,\yy}\label{67}
\ee
%
%
%
%
with $\int d\xx=a^2\sum_\xx$ and 
$P(d\psi)$ and $P(d\phi)$ are the fermionic and bosonic integration with propagator $\d_{\a,\a'} g(\xx)$ with
\be g(\xx)
={1\over L^2}\sum_{\kk}e^{i\kk\xx}\hat\chi(\kk)\begin{pmatrix} {1\over a}
[i\sin a k_1-\sin a k_2]& -E \\ -E & {1\over a}[i\sin a k_1+\sin a k_2]  
\end{pmatrix}^{-1}\label{pro}
\ee
%
%
%
%
%
%
and  we have used that the normalization of the bosonic and fermionic
integration are one inverse to the other. 
The fermionic sector of the above functional integral coincides with a massless Thirring model with a {\it non local} current-current interaction.

In the same way we can rewrite the averaged product of two functions as
%
%
%
\bea
&&K_{\l,E,N}(\xx)=
\mathbb{E}\{\int \DD\Psi_a e^{-(\psi_a^+,A \psi^-_a)-(\phi_a^+,A \phi^-_a)- }
\psi^-_{a,\o,\xx}\psi^+_{a,\o,0}\nn\\
&&
\int \DD\Psi_b e^{-(\psi_b^+, B\psi_b^-) -(\phi_b^+, B\phi^-_b)}
\psi^-_{b,\o,\xx}\psi^+_{b,\o,0}\}\label{89}
\eea
with $A=i H- E$, $B=i H+ E$, 
and averaging over the disorder
%
%
%
%
\be
K_{\l,E,N}(\xx)=\int P(d\Psi_a) P(d\Psi_b)\e^{\VV}
\psi^-_{a,\o,\xx}\psi^+_{a,\o,0}
\psi^-_{b,\o,\xx}\psi^+_{b,\o,0}]\label{24}
\ee
where 
\be\VV=
-\l\int d\xx \int d\yy v(\xx-\yy)\sum_{\a,\a=\phi,\psi\atop \b,\b'=a,b}\sum_{\o=\pm}\Psi^+_{\a,\b,\o,\xx}\Psi_{\a,\b,\o,\xx}\Psi^+_{\a',\b',-\o,\yy}\Psi_{\a',\b',-\o,\yy}
\ee

\section{The critical theory}

\subsection{The averaged two point function}

We define the  {\it generating function} as
\be
e^{\WW_N(J,\h)}=
\int P(d\Psi)  e^{\VV(\Psi)+\int d\xx [(\h^-_\xx,\Psi^+_{\xx})+
 (\h^+_\xx,\Psi^-_{\xx})+(J_\xx,\r_\xx)]}\label{al}
\ee
where $\r_{\o,\a,\xx}=\Psi^+_{\o,\a,\xx}\Psi^-_{\o,\a,\xx}$
and we define,  for $\a=(\psi,\phi)$, $\o=\pm$, the truncated correlations 
\bea
&&<\Psi^-_{\a,\o,\xx}\Psi^+_{\a,\o,\yy}>_{T,E,N}
={\partial^2\over \partial \h^+_{\a,\o,\xx}\partial\h^-_{\a,\o,\yy}}\WW_N(J,\h)|_{0}
\label{zaz}  \\
&&<\r_{\a',\o',\zz}\Psi^-_{\a,\o\xx}\Psi^+_{\a,\o\yy}>_{T,E,N}
={\partial^3\over \partial J_{\a',\o',\zz}\partial \h^+_{\a,\o,\xx}\partial\h^-_{\a,\o,\yy}}\WW_N(J,\h)|_{0}\nn\\
\eea
where $<AB>_T=<AB>-<A><B>$ and 
$<\Psi^-_{\psi,\o,\xx}\Psi^+_{\psi,\o',0}>_{T,E,N}\equiv <\psi^-_{\o,\xx}\psi^+_{\o',0}>_{T,E,N}
\equiv G_{\l,E,N;\o,\o'}(\xx)$ defined by \pref{88}.

Using a smooth decomposition of the unity, 
we write the propagator as sum of single scale propagators at $E=0$
\be
g(\xx)=\sum_{j=h_L}^N g^{(j)}(\xx)
\ee
where $h_L\sim -\log L$ and $g^{(j)}(\xx)$, the single scale propagator, is similar to 
$g(\xx)$ \pref{pro} with $\hat\chi(\kk)$ replaced by $f_j(\kk)$ , with 
$f_j(\kk)$ non vanishing in $2^{j-1}\le |\kk|\le 2^{j+1}$. The presence of a minimal scale $h_L$ comes from the fact that antiperiodic boundary conditions are assumed,
and therefore the momenta are of the form $\kk={2\pi\over L}({\bf n}+{1\over 2})$, so that $|k_i|\ge {\pi\over L}$. $L$ plays the role of an {\it infrared cut-off} while $2^N$ is the {\it ultraviolet} cut-off.
%
%
%
%
%
Note that
\bea 
&&|g^{(j)}|_{L_1}=\int d\xx |g^{(j)}(\xx)|\le C  2^{-j}\nn\\
&&|g^{(k)}|_{L_\io}=\sup_\xx |g^{(j)}(\xx)|
\le C
2^{j} \label{fon111}
\eea
We use now the following basic property of  gaussian integrations, bosonic or fermionic, called  {\it addition
property} and we get
%
%
, calling
the exponent in the r.h.s. of \pref{al} simply $V(\Psi,\h,J)$
\be \int P(d\Psi)e^{V(\Psi,\h,J) }=\int P(d\Psi^{(\le N-1)}) \int
P(d\Psi^{(N)}) e^{V(\Psi,\h,J)} =\int P(d\Psi^{(\le
N-1)})e^{V^{(N-1)}(\Psi^{(\le N-1)}\h,J)} \label{all} \ee
where $P(d\Psi^{(\le N-1)})$ and $P(d\Psi^{(N)})$ are gaussian
integrations with propagator respectively $g^{(\le N-1)}(\xx)$ and  $g^{(\le N)}(\xx)$
and
\be V^{(N-1)}(\Psi,\h,J)=\sum_{n=1}^\io {1\over n!} \EE^T(V;n)\ee
with $\EE^T$ are the {\it truncated expectations}  with respect to
$P(d\Psi^{(N)})$
\be
\EE^T(V(\Phi);n)={\partial^n\over \partial \a^n}\log \int P(d\Psi^{(N)}) e^{\a V(\Psi^{(N)}+\Phi)}|_{\a=0}
\ee
When expressed in terms of Feynman graphs, the truncated expectation are written in terms of {\it connected} diagrams only.
The multiscale analysis continues integrating the
fields $\Psi^{(N-1)},..,\Psi^{(h+1)}$  obtaining
\be e^{\WW_N(J,\h)}
=\int P(d\Psi^{(\le h)})e^{V^{(h)}(\Psi^{(\le h)},\h,J)}\ee
with $V^{(h)}$, called effective potential, being a sum of integral of monomials with $n\ge 0$
$\Psi,\h$ and $m\ge 0$ $J$ fields multiplied by
kernels $W_{n,m}^{(h)}$; moreover $P(d\Psi^{(\le h)})$ is the
integration with propagator $g^{(\le h)}(\xx)= \sum_{k \le h}
g^{(k)}(\xx)$. 

The range of the disorder ($\k=1$ in \pref{alal}) provides a  natural {\it momentum scale}
separating the scales $j$ in {\it ultraviolet scales}, between $0$ and $N$, and {\it infrared scales}, namely between $h_L$ and $-1$.  Let us consider first the integration of the ultraviolet scales.
The {\it scaling dimension} in the case of $\d$-correlated disorder is the same in the ultraviolet and infrared
region and equal to  
$D=2-n/2-m$, that is greater or equal to zero in the case $n,m=(2,0), (4,0), (2,1)$.
Therefore there are in general ultraviolet divergences and this requires that
the ultraviolet $N\to\io$ limit can be taken only choosing properly the bare parameters to  $N$-dependent and possibly singular value.
For instance, in the case of the Thirring model with a local $\d$-like interaction, 
the $N\to\io$ limit can be taken only choosing the bare wave function renormalization vanishing as $2^{-\h N}$
with $\h>0$. 

In the case of short-ranged correlated disorder the situation is different; the non locality of the disorder
induces an {\it improvement} in the scaling dimension, and 
indeed no ultraviolet divergences are present; the kernels 
of the effective potential are bounded uniformly in the ultraviolet cut-off $N$.
Consider for instance $W^{(h)}_{2,0}$, $h\ge 0$, with scaling dimension $D=1$. We can decompose
$W^{(h)}_{2,0}$, using general properties of truncated expectations (or the fact that they are expressed in terms of connected diagrams),
as in Fig. 1. 
\insertplot{500}{100} {\ins{130pt}{60pt}{$+$}
\ins{50pt}{60pt}{$=$}
\ins{280pt}{60pt}{$+$}
} {verticiT11}
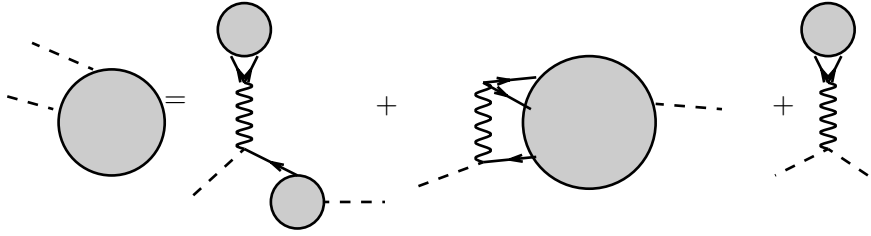
{\label{h2} Graphical representation of the decomposition of the
kernel $W_{2,0}^{(h)}$; ; the blobs represent $W^{(h)}_{n,m}$, the
paired wiggly lines represent $v$, the full lines $g^{(h,N)}$ and
the dotted lines are the external fields} {0}
Note that the first and third contributions
are vanishing by parity considerations (remember that $E=0$ here); regarding the second, we can use the following bound 
\bea &&|\int d\xx_1 d\xx_2 d\xx_3 v(\xx_1-\xx_2)
g^{[h,N]}(\xx_1-\xx_3) W_{2,1}^{(h)}(\xx_2;\xx_3,0)|\le\nn\\
&&|g^{[h,N]}|_{L_1} |v|_{L_\io} \int d\xx_2 d\xx_3
|W_{2,1}^{(h)}(\xx_2;\xx_3,0)|\le C 2^{-h}, \eea
where we have inductively bounded 
$|W_{2,1}^{(h)}|_{L_1}$
by a constant, as its dimension is $D=0$. 
Note the crucial role played by the non locality of the disorder; in the case of
$\d$-correlated disorder one needs to integrate over the wiggly lines instead than over the propagator
(as 
$|v|_{L_\io} $ is unbounded) so that in the above bound one gets 
$|g^{[h,N]}|_{L_\io} |v|_{L_1}$ instead of 
$|g^{[h,N]}|_{L_1} |v|_{L_\io}$ 
and the resulting bound would be diverging as $N\to\io$ as $2^N$.
Similar considerations could be done for $W_{0, 2}^{(h)}$ which can be
decomposed as in the r.h.s. Fig. 2; the second term can
again be bounded by
\bea && |v|_{L^\io} |W^{(k)}_{2,2}|_{L^1}\sum_{h\le i'\le
j\le i\le N}|g^{(j)}|_{L^1} |g^{(i)}|_{L^1}|g^{(i')}|_{L^\io}\le\nn\\
&& C_1 \l 2^{-2h}\sum_{h\le i\le N} (i-h)2^{-i+h}\le C_2
\l 2^{-2h}\label{aza} \eea
\insertplot{500}{100} {\ins{150pt}{60pt}{$+$}
} {verticiT15x}
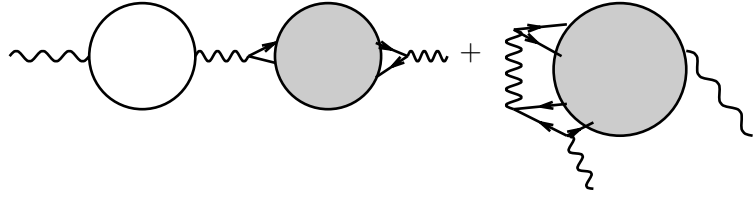
{\label{h4} Decomposition of $W^{(k)}_{0,2}$: the blobs represent
$W^{(k)}_{n,m}$, the paired wiggly lines represent $v$, the paired
line $g^{(k,N)}$ }{0}
This argument again cannot be repeated for the first term in Fig. 2, but 
the local part vanishes since the local part of the bubble graph is zero
by symmetry
\be
{1\over L^2}\sum_\kk \chi_{N}(\kk){k_0^2- k^2+2 i  k_0
k\over (k_0^2+ k^2)^2}= 0. \label{xxx} \ee
A similar analysis can be repeated for the other terms to show
that the scaling dimension is always negative. The conclusion is that 
the effective potential is uniformly bounded in $N$ and that
there are 
{\it no ultraviolet divergences} even when the momentum cut-off is removed, that is for $N\to\io$.

\subsection{Ward Identities and cancellation of the anomalies}

A crucial role is played by {\it Ward Identities}, which can be obtained 
by performing in \pref{al} with $E=0$ the {\it chiral} local phase transformation
$
\Psi^\pm_{\o,\a,\xx}\to e^{\pm i a_{\o,\a,\xx}} \Psi^\pm_{\o,\a,\xx}
$ and performing a derivative with respect to $a_{\o,\a,\xx}$ and the external fields;
due to the presence of cut-offs the Jacobian is 
equal to $1$ but, with respect to the formal Ward Identities obtained neglecting cut-offs, one has an extra term; indeed it is found
\bea
&&D_\o(\pp)<\hat\r_{\a,\o,\pp}\hat\Psi^-_{\a',\o'\kk}\hat\Psi^+_{\a',\o',\kk+\pp}>_{T,0,N}=
\d_{\a,\a'}\d_{\o,\o'}[<\hat\Psi^-_{\a',\o',\kk}\hat\Psi^+_{\a',\o',\kk}>_{T,0,N}\nn\\&&
-<\hat\Psi^-_{\a',\o',\kk+\pp}\hat\Psi^+_{\a',\o',\kk+\pp}>_{T,0,N}]
+<\d \hat\r_{\a,\o,\pp}\hat\Psi^-_{\a',\o',\kk}\hat\Psi^+_{\a',\o',\kk+\pp}>_{T,0,N}\label{pp}
\eea
where $D_\o(\kk)=i k_1-\o k_2$, with  $\o=\pm $,
$\r_{\a,\o,\xx}=\Psi^+_{\a,\o,\xx}\Psi^-_{\a,\o,\xx}$,
\be
\d\hat\r_{\a,\o,\pp}=\int d\kk C_{N}(\kk,\pp)\hat\Psi^+_{\a,\o,\kk}\hat\Psi^-_{\a,\o,\kk+\pp}
\ee
with
\be C_{N}(\kk,\pp)=[\hat\chi(\kk + \pp)^{-1} - 1]
D_{\o}(\kk+\qq) - [ (\hat\chi(\kk)^{-1} - 1]
D_{\o}(\kk)\label{fonddd1}. \ee
The last term in \pref{pp} is due the the presence of the momentum cut-off which breaks the {\it local} chiral invariance. 
Remarkably, such term it is not vanishing even removing the ultraviolet cut-off, 
but the following identity holds
\bea
&&\
<\d \hat\r_{\a,\o,\pp}\hat\Psi^-_{\a',\o',\kk}\hat\Psi^+_{\a',\o,\kk+\pp}>_{T,0,N}=\nn\\
&&
\e_\a\l {1\over 4\pi} D_{-\o}(\pp)\sum_{\a''=\phi,\psi}<\hat\r_{\a'',-\o,\pp}\hat\Psi^-_{\a',\o',\kk}\hat\Psi^+_{\a',\o',\kk+\pp}>_{T, 0,N}+R_{N,\a}(\kk,\pp)\label{aaaa7}
\eea
with
\be
\e_\psi=-1,\quad\quad\quad \e_{\phi}=1\label{aaadd}
\ee
and $R_{N,\a}(\kk,\pp)$ is in absolute value smaller than ${
2^{-N}\over |\kk||\kk-\pp|}$, that is vanishing for $N\to\io$. 

If we restrict to the fermionic sector, in the limit $N\to\io$
the first term in the r.h.s. of (32)
would be the chiral anomaly and the second term is vanishing. 
The fact that the chiral anomaly is {\it linear}
in the coupling is a property called 
Adler-Bardeen theorem. It is important to stress the presence of the correction term $R_N$ 
in the l.h.s. of (32),
which is vanishing {\it only} in the $N\to\io$ limit.
The derivation of (32) is based on a 
multiscale integration also for the correction
term in \pref{pp}, and the main difference is that the source term
$(J_\xx,\Psi^+_{\xx}\Psi_\xx)$ is replaced by $\int d\xx \chi_{\a,\o}\d\r_{\a,\o,\xx}$
where $\chi$ is a source term.
After the integration of the fields
$\Psi^{(N)},\Psi^{(N-1)},..,\Psi^{(h+1)}$ the effective potential can be again written as sum of monomials with $n$ $\Psi$ fields, and $m$ $\chi$ fields with
kernels $\tilde W^{(h)}_{n,m}$.
\insertplot{500}{130} {\ins{150pt}{60pt}{$+$}
} {verticiT15xx}
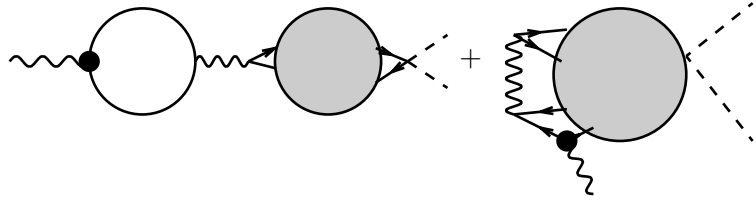
{\label{m9} Contributions to $\tilde W^{(h)}_{1,2}$; the black dot
represents the $\chi \d \r$ vertex} {0}
The analysis of  $\tilde W^{(h)}_{2,1}$ is very similar to the analysis of 
 $W^{(h)}_{2,1}$ in the previous section. An important difference with respect to the bound 
\pref{aza} comes from 
the fact that
$C_{N}(\kk,\pp)
g^{(i)}(\kk)g^{(j)}(\kk+\pp)$
vanishes unless either $i$ or $j$ equals the cut-offs
scales $N$. Therefore, the second term in
Fig. 3, which contributes to $R_{N,\a}$ in (32), can be bounded as \pref{aza} with the difference that one of the scales of the propagator 
attached to the back dot have scale $N$; therefore one obtains the bound
\bea &&|\l| |v|_{L^\io} | W^{(h)}_{4,1}|_{L^1}\sum_{h\le
i'\le i\le N}|g^{(N)}|_{L^1}
|g^{(i)}|_{L^1}|g^{(i')}|_{L^\io}\le\label{xaxa}\\
&&C_1\l^2 2^{-2h}(N-h)2^{-N+h}\le C_2\l^2
2^{-2h}2^{-(N-h)/2}\nn\eea
leading to the vanishing of this contribution for $N\to\io$.
On the other hand the non-connected contributions, that is the first term in Fig. 3,
 is now non vanishing and contribute to the first term in (32) ;
the bubble is now given by
\be \e_\a{1\over L^2}\sum_{\kk}{C_{N}(\kk,\pp)\over
D_{-\o}(\pp)} g_\o(\kk)g_\o(\kk+\pp)=\e_\a{1\over 4 \pi}+O(2^{-N})
\ee
%
%
%

Remarkably, the anomalies cancel out in the WI for the total density due to supersymmetry (that is, due to \pref{aaadd})
\bea
&&D_\o(\pp)\sum_{\a=\phi,\psi}<\hat\r_{\a,\o,\pp}\hat\Psi^-_{\a',\o',\kk}\hat\Psi^+_{\a',\o',\kk+\pp}>_{T,0,N}=\d_{\o,\o'}
[<\hat\Psi^-_{\a',\o',\kk}\hat\Psi^+_{\a',\o',\kk}>_{T,0,N}\nn\\&&
-<\hat\Psi^-_{\a',\o',\kk+\pp}\hat\Psi^+_{\a',
\o',\kk+\pp}>_{T,0,N}]
+\sum_{\a=\phi,\psi} R_{N,\a}(\kk,\pp)\label{pp1}
\eea
We can write the Schwinger-Dyson equation
\be
<\hat\Psi^-_{\a,\o,\kk}\hat\Psi^+_{\a,\o,\kk}>_{T,0,N}=\hat g_\o(\kk)+\l \hat g_{\o}(\kk)
\int d\pp \hat v(\pp) \sum_{\a'=\phi,\psi}<\hat\r_{\a',-\o}(\pp)\Psi^+_{\a,\o,\kk+\pp}\Psi^-_{\a,\o,\kk}>_{T,0,N}
\label{sd}\ee
and inserting \pref{pp1} we obtain 
\be
<\hat\Psi^-_{\a,\o,\kk}\hat\Psi^+_{\a,\o,\kk}>_{T,0,N}=\hat g_\o(\kk)+\l \hat g_\o(\kk)
\int d\pp \hat v(\pp) \sum_{\a'=\phi,\psi} {R_{N,\a'}(\kk,\pp)\over D_\o(\pp)}\label{sd1}
\ee
It can be shown, by an analysis similar to the one for $R_{N,\a}$ \pref{xaxa}, that 
 that $\int d\pp \hat v \sum_{\b=\phi,\psi} {R_{N,\a}\over D_\o}$
is smaller than $2^{-N}$, that is vanishing when the ultraviolet cut-off is removed. 
Therefore at $E=0$ the averaged 2-point function is equal to the free one up to corrections 
which are vanishing {\it only} when the ultraviolet cut-off is removed $N\to\io$; on the other hand for any finite
cut-off non vanishing corrections are expected.

It is indeed useful to compare the present result to 
the analogous computation for the Thirring model with non local interaction, that is neglecting the bosons; in such a case
the Schwinger-Dyson equation is still given by \pref{sd} but in the r.h.s. 
$\sum_{\a=\phi,\psi}
\hat\r_{\a,-\o,\a}(\pp)$ should be replaced by 
$\hat\r_{\psi,-\o}(\pp)$; by using the WI \pref{pp}, (32) one would get an extra term in 
\pref{sd1} function of the chiral anomaly. As a result, one would find that the asymptotic behavior of $<\hat\Psi^-_{\psi,\o,\kk}\hat\Psi^+_{\psi,\o,\kk}>_{T,0,N}$ is {\it different} with respect to the non interacting case;
for small $\kk$
$<\hat\Psi^-_{\psi,\o,\kk}
\hat\Psi^+_{\psi,\o,\kk}>_{T,0,N}$ would behave as $|\kk|^{-1+\h}$ with $\h>0$. In the present case, instead, the cancellation
of the anomalies due to the supersymmetry has the effect that the asymptotic behavior  of the two point function is equal to the free one, up to a small correction vanishing  when the cut-off is removed.

\subsection{The average of the product}

Starting from \pref{24} for $E=0$, and using the notation \pref{zaz}
(with $P(d\Psi)$ replaced by 
$P(d\Psi_a) P(d\Psi_b)$) we can write
\be
K_{\l,0}(\xx)=
<\psi^-_{a,\o,\xx}\psi^+_{a,\o,0}>_{T,0,N}
<\psi^-_{b,\o,\xx}\psi^+_{b,\o,0}>_{T,0,N}+<\psi^-_{a,\o,\xx}\psi^+_{a,\o,0}
\psi^-_{b,\o,\xx}\psi^+_{b,\o,0}>_{T,0,N}\label{ma1}
\ee
The computation of $<\psi^-_{a,\o,\xx}\psi^+_{a,\o,\yy}>_{T,0,N}$ can be done exactly as in the previous case. The Schwinger-Dyson equation is given by
\be
<\hat\psi^-_{\b,\o,\kk}\hat\psi^+_{\b,\o,\kk}>_{T,0,N}=\hat g_\o(\kk)+\l \hat g_\o(\kk)
\int d\pp \hat v(\pp) \sum_{\a'=\phi,\psi}\sum_{\b'=a,b}<\hat\r_{\a',\b',-\o}(\pp)\hat\psi^+_{\b,\o,\kk+\pp}\hat\psi^-_{\b,\o,\kk}>_{T,0,N}
\ee
and using the WI
\bea
&&D_\o(\pp)<\hat\r_{\a,\b,\o,\pp}\hat\psi^-_{\b',\o',\kk}\hat\psi^+_{\b',\o',\kk+\pp}>_{T,0,N}=
\d_{\a,\psi}\d_{\b,\b'}\d_{\o,\o'}[<\hat\psi^-_{\b',\o',\kk}\hat\psi^+_{\b',\o',\kk}>_{T,0,N}\nn\\&&
-<\hat\psi^-_{\b',\o',\kk+\pp}\hat\psi^+_{\b',\o',\kk+\pp}>]
+<\d \hat\r_{\a,\b,\o,\pp}\hat\psi^-_{\b',\o',\kk}\hat\psi^+_{\b',\o',\kk+\pp}>_{T,0,N}\label{pp3}
\eea
where $\r_{\a,\b,\o,\xx}=\Psi^+_{\a,\b,\o,\xx}\Psi^-_{\a,\b,\o,\xx}$ and again
\bea
&&
<\d \hat\r_{\a,\b,\o,\pp}\hat\psi^-_{\b',\o',\kk}\hat\psi^+_{\b',\o',\kk+\pp}>_{T,0,N}=\\
&&
\e_\a\l {1\over 4\pi} D_{-\o}(\pp)\sum_{\a''=\phi,\psi\atop \b''=a,b}
<\hat\r_{\a'',\b'',-\o,\pp}\hat\psi^-_{\b',\o',\kk}\hat\psi^+_{\b',\o',\kk+\pp}>_{T,0,N}+R^{(2)}_{N,\a,\b}(\kk,\pp)\label{aaa1}\nn
\eea
we get
\be
<\hat\psi^-_{\b,\o,\kk}\hat\psi^+_{\b,\o,\kk}>_{T,0,N}=\hat g_\o(\kk)+\l \hat g_\o(\kk)
\int d\pp \hat v(\pp) \sum_{\a'=\phi,\psi,\b'=a,b} {R^{(2)}_{N,\a',\b'}(\kk,\pp)\over D_\o(\pp)}\label{ma2}
\ee
with $\int d\pp \hat v(\pp) {R^{(2)}_{N,\a',\b'}\over D_\o}$
is $O(2^{-N})$.
In a similar way one analyze the connected part of \pref{ma1}; we write 
the Schwinger-Dyson equation for the four point function
\be
<\hat \psi^-_{a,\o,\kk_1}\hat\psi^+_{a,\o,\kk_2}\hat\psi^-_{b,\o,\kk_3}\hat\psi^+_{b,\o,\kk_4}>_{T,0,N}=
\int d\pp \hat v(\pp) 
\sum_{\a'=\phi,\psi\atop \b'=a,b} <\hat\r_{\a',\b',-\o,\pp}
\hat\psi^-_{a,\o,\kk_1}\hat\psi^+_{a,\o,\kk_2}\hat\psi^-_{b,\o,\kk_3}\hat\psi^+_{b,\o,\kk_4-\pp}>_{T,0,N}\ee
The WI for the four point function is
\bea
&&D_\o(\pp)
<\hat\r_{\a',\b',-\o,\pp}
\hat\psi^-_{a,\o,\kk_1}\hat\psi^+_{a,\o,\kk_2}\hat\psi^-_{b,\o,\kk_3}\hat\psi^+_{b,\o,\kk_4-\pp}>_{T,0,N}
+\nn\\
&&<\d \hat\r_{\a',\b',-\o,\pp}
\hat\psi^-_{a,\o,\kk_1}\hat\psi^+_{a,\o,\kk_2}\hat\psi^-_{b,\o,\kk_3}\hat\psi^+_{b,\o,\kk_4-\pp}>_{T,0,N}
=0
\eea
with 
\bea
&&
<\d \hat\r_{\a',\b',-\o,\pp}\hat\psi^-_{a,\o,\kk_1}\hat\psi^+_{a,\o,\kk_2}\hat\psi^-_{b,\o,\kk_3}\hat\psi^+_{b,\o,\kk_4-\pp}>_{T,0,N}=\\
&&
\e_{\a'}\l {1\over 4\pi} D_{\o}(\pp)\sum_{\a'',\b''}<\hat\r_{\a'',\b'',\o,\pp}\hat\psi^-_{a,\o,\kk_1}\hat\psi^+_{a,\o,\kk_2}\hat\psi^-_{b,\o,\kk_3}\hat\psi^+_{b,\o,\kk_4-\pp}>_{T,0,N}+R^{(4)}_{N,\a',\b'}(\kk_1,\kk_2,\kk_3,\kk_4,\pp)\label{aaa}\nn
\eea
and using that  $\sum_{\a'=\phi,\psi} \e_{\a'}=0$ we finally obtain
\be
<\hat \psi^-_{a,\o,\kk_1}\hat\psi^+_{a,\o,\kk_2}\hat\psi^-_{b,\o,\kk_3}\hat\psi^+_{b,\o,\kk_4}>_{T,0,N}=
\sum_{\a',\b'}\int d\pp \hat v(\pp) {R^{(4)}_{N,\a',\b'}\over D_{\o}(\pp)}\label{ma3}
\ee
and again the r.h.s. is vanishing as $O(\l 2^{-N}))$. Therefore, by \pref{ma1},\pref{ma2},\pref{ma3}
the interacting average of the product 
$K_{\l,0,N}(\xx)$ \pref{89} differs from its non interacting value $K_{0,0,N}(\xx)$ by terms which are order $O(\l 2^{-N})$ for large $N$; exact universality for such quantity (and therefore for the conductivity) is achieved only in the limit of removed ultraviolet cut-off.

\section{The non critical theory and the infinite volume limit}

We have to discuss finally the removal of the infrared cut-off and the case $E\not =0$.
Again we perform a multiscale decomposition of $\Psi$ in the $E\not=0$ case and  the integration of the ultraviolet (positive) scales 
is done as in the previous section (the fact that $E\not=0$ plays no role in the ultraviolet regime).
We consider now
the integration of the negative infrared scales, in the $L\to\io$ limit. In this case there is no improvement with respect to the scaling dimension, and one has to define
a {\it renormalized} multiscale integration in the following way.
Assume that we have integrated the fields
$\Psi^{(N)},..,\Psi^{(h)}$, $h\le 0$  obtaining
\be e^{\WW(J,\phi)}
=\int P_{Z_h,E_h}(d\Psi^{(\le h)})e^{V^{(h)}(\sqrt{Z_h}\Psi,\h,J)}\ee
where  $P_{Z_h,E_h}(d\Psi^{(\le h)})$ is the gaussian integration with propagator, $\a=\psi,\phi$
\be
g_{\a}(\xx,\yy) ={1\over L^2}
\sum_{\kk}\hat\chi_h(\kk)
e^{i\kk(\xx-\yy)} {1\over Z^{(\a)}_h}\begin{pmatrix} D_+(\kk) & E^{(\a)}_h \\  E^{(\a)}_h & D_-(\kk)
\end{pmatrix}^{-1}
\ee
and $\hat\chi_h(\kk)=\sum_{j=-\io}^h f_j(\kk)$ and again
$V^{(h)}$ being a sum of integral of monomials with $n\ge 0$
$\Psi,\h$ and $m\ge 0$ $J$ fields multiplied by
kernels $\hat W_{n,m}^{(h)}$.
We decompose the kernels
as
\be
\hat W_{n,m}^{(h)}(\underline \kk)=\hat W_{n,m;a}^{(h)}(
\underline\kk)+\hat W_{n,m;b}^{(h)}(
\underline\kk)+\hat W_{n,m;c}^{(h)}(
\underline\kk)
\ee
where $\hat W_{n,m;a}^{(h)}$ and $\hat W_{n,m;b}^{(h)}$ are respectively
the zero-th and first order contribution in $E$ to
$\hat W_{n,m}^{(h)}$ and $\hat W_{n,m;c}^{(h)}$ is the rest.  We define a {\it localization operator}
on the terms with positive scaling dimension $D=2-n/2-m$
in the following way
\bea
&&\LL  \hat W^{(h)}_{4,0}(\kk_1,\kk_2,\kk_3,\kk_4)=\hat W^{(h)}_{4,0;a}({\bf 0}, {\bf 0},{\bf 0},{\bf 0})\nn\\
&&\LL \hat W^{(h)}_{2,1}(\kk_1,\kk_2,\kk_3)=\hat W^{(h)}_{2,1;a}({\bf 0},{\bf 0},{\bf 0})\nn\\
&&\LL  \hat W^{(h)}_{2,0;\o,\o}(\kk)= \hat W^{(h)}_{2,0:\o,\o;a}({\bf 0})+\kk {\bf \partial} \hat W^{(h)}_{2,0;\o,\o;a}({\bf 0})\nn\\
&&\LL\hat  W^{(h)}_{2,0;\o,-\o}(\kk)=\hat W^{(h)}_{2,0:\o,-\o;a}({\bf 0})+ \hat W^{(h)}_{2,0:\o,-\o;b}({\bf 0})
\eea
and we write
\be e^{\WW(J,\phi)}
=\int P_{Z_h,E_h}(d\Psi^{(\le h)})e^{\LL V^{(h)}(\Psi,\h,J)+(1-\LL)V^{(h)}(\sqrt{Z_{h-1}}\Psi,\h,J) }\label{fg}\ee
The action of $(1-\LL)$ on the kernels improve their scaling dimension. For instance
\be
(1-\LL)\hat W^{(h)}_{4,0}=
[\hat W^{(h)}_{4,0;a}(\underline\kk)- \hat W^{(h)}_{4,0;a}(\underline {\bf 0})]+\hat W^{(h)}_{4,0;b}(\underline\kk)\label{41}
\ee
and the first term in the r.h.s. has negative dimension while
%
%
regarding 
the other term one has simply to use that
%
%
the bound for $\hat W^{(h)}_{4,0;a}$ as an extra ${E_h^{(\a)}\over 2^h}$.

We use now the following symmetries of the propagator at $E=0$
\be
(g^{(k)}_\o)^*(k_1,k_2)=g_{\o}^{(h)}(-k_1,k_2)\quad\quad g^{(h)}_\o(k_1,k_2)=-i\o g^{(h)}_{\o}(k_2,-k_1)\label{bb1}
\ee
and that at $E=0$ there is global phase invariance
$
\Psi^\pm_{\a,\o}\to e^{\pm i \a_{\a,\o}}\Psi^\pm_{\a,\o}
$.
%
%

Therefore
\begin{enumerate}
\item The local part of the terms with four fields with the same $\o$ is vanishing; indeed if $n$ is the order there are
$n-2$ $(\o)$-propagators and $n$ $(-\o)$-propagators; then by \pref{bb1} $\hat W^{(h)}_{4,0;a}(\underline k_1,\underline k_2)=(i\o)^{-2} W^{(h)}_{4,0}(-\underline k_2,\underline k_1)$ so that $\hat W^{(h)}_{4,0;a}
(\underline 0,\underline 0)=-\hat W^{(h)}_{4,0;a}(\underline 0,\underline 0)=0$; moreover by global phase invariance there is an even number
of fields with the same $(\a,\o)$.
\item The quartic terms are real. Indeed by \pref{bb1} $(\hat W^{(h)}_{4,0;a})^*(\underline k_1,\underline k_2)=
\hat W^{(h)}_{4,0;a}(-\underline k_1,\underline k_2)$, so that the local part is real
\item The local part of the terms with two external line and the same $\o$ is vanishing by the parity of the
propagator.
\item
Finally $\partial_1 \hat W^{(h)}_{2,0;a}(0)=i\o \partial_2  \hat W^{(h)}_{2,0:a}(0)$
\end{enumerate}
The only quadratic terms in $\LL\VV^{(h)}$ are the one multiplying $\partial W^{(h)}_{2,0;\e,\e;a}({\bf 0})$ and
$ W^{(h)}_{2,0:\e,-\e}({\bf 0})$ producing respectively a renormalization of $Z^{(\a)}_h$ and $E^{(\a)}_h$. Therefore we can move
the quadratic terms in the gaussian integration so obtaining
\be
\int P_{Z_{h-1,}E_{h-1}}(d\Psi^{(\le h)})e^{\LL \tilde V^{(h)}(\sqrt{Z_{h-1}}\Psi,\h,J)+(1-\LL)V^{(h)}(\sqrt{Z_{h-1}}\Psi,\h,J) }\label{fga}\ee
and
\bea
&&\LL \tilde V^{(h)}(\Psi,0,0)=
\l_{1,h}\sum_{\o}\int d\xx \psi^+_{\xx,\o}\psi^-_{\xx,\o}
\psi^+_{\xx,-\o}\psi^-_{\xx,-\o}+\\
&&
\l_{2,h}\sum_{\o}\int d\xx \phi^+_{\xx,\o}\phi^-_{\xx,\o}
\phi^+_{\xx,-\o}\phi^-_{\xx,-\o}+
\l_{3,h}\sum_{\o}\int d\xx \psi^+_{\xx,\o}\psi^-_{\xx,\o}
\phi^+_{\xx,-\o}\phi^-_{\xx,-\o}\nn
\eea
One can write \pref{fga} as
\be
\int P_{Z_{h-1,}E_{h-1}}(d\Psi^{(\le h-1)})\int  P_{Z_{h-1,}E_{h-1}}(d\Psi^{(h)}) e^{\LL \tilde V^{(h)}(\sqrt{Z_{h-1}}\Psi,\h,J)+(1-\LL)V^{(h)}(\Psi,\h,J) }\label{fgaa}\ee
and the procedure can be iterated up to a scale $h^*_\a$ (that is
 $h^*_\psi$ for the fermionic fields and  $h^*_\phi$ for the bosonic ones)
 such that
$E_{h^*_\a}=2^{h^*_\a}$; one can see that $g^{(-\io, h^*)}$ obey exactly to the same bounds as the single scale propagator $g^{(h)}$ with $h>h^*_\a$.
The outcome of this procedure is a sequence of
$V^{(h)}(\Psi,\h,J)$ with kernels $W^{(h)}_{n,m}$, expressed in terms of 
the effective coupling constants
$\l_{i,k}$, $k=h,h+1,..0$; the kernels are finite uniformly in $h$ provided that 
the running coupling constants stay bounded. On the other hand the running coupling constants are the same
in the critical theory at $E=0$. Therefore in order to study their flow 
can consider the theory with $E=0$ and infrared cut-off $2^h$, replacing $\hat\chi(\kk)$ with
$\hat\chi_{h,N}(\kk)=\sum_{j=h}^{N} f_j(\kk)$
with $h\le 0$. The 
Schwinger-Dyson equations for the two and four point function coincide with the ones derived in the previous sections up to negligible corrections
due to the presence of the infrared cut-off $2^h$.
Therefore fixing the value of the external momenta at the scale of the infrared cut-off we get
\be\l_{h-1,i}=\l_{0,i}+O(\l_0^2)
\quad\quad
Z_h^{(\a)}=1+O(\l_0^2)\label{x1}
\ee
and $\l_0=\l\hat v(0)+O(\l^2)$. This means that the effective couplings $\l_{h,i}$ converge to a line of fixed points (the beta function is asymptotically vanishing)
 and the critical exponent for the wave function renormalization is zero (contrary to what happens in 
the fermionic theory in which is positive).

The flow equation for the energy is given by
\be
{E^{(\psi)}_{h-1}\over E^{(\psi)}_h}=1+a\l_{1,h}+O(\l_h^2)\quad\quad  {E^{(\phi)}_{h-1}\over E^{(\phi)}_h}=1+a\l_{2,h}+O(\l_h^2)
\ee
with $a={1\over 2\pi}>0$ and by symmetry the contributions with different $\a$ do not mix, by the global phase symmetry valid at $E=0$.
Therefore
\be
E^{(\a)}_h= E 2^{-\h_\a h}
\ee
with $\h_\a=a\hat v(0)\l+O(\l^2)$; this implies $2^{h^*_\a}=E^{1\over 1+\h_\a }$.
For $h\ge h^*=\max (h^*_\phi, h^*_\psi)$ this makes clear why
the second term in \pref{41} has the correct scaling dimension;
indeed $
E^{(\a)}_h 2^{-h}$ can be bounded by $2^{(1+\h_\a)(h^*-h)}$
which is sufficient to make the dimension negative. For $h\le h^*$ the theory becomes purely fermionic or bosonic.
Therefore
\be
<\hat\psi^-_{\kk,\o}\hat\psi^+_{\kk,-\o}>
=
\sum_{h=h^*_\psi}^\io
{E^{(\psi)}_h\over Z^{(\psi)}_h}{f_h(\kk)\over k_0^2+k^2}(1+\l F_h)
\ee
with $|F_h(\kk)|\le \l$ and $E^{(\psi)}_h=E$, $Z_h=1$ f
or $h\ge 0$; we have used that the contributions from the scales $h \le h^*$ are summable.
The density of states (with imaginary energy) is therefore bounded by
\be
\sum_{h=h^*_\psi}^0 |E^{(\psi)}_h|+E\sum_{h=0}^\io e^{-\e 2^h} \le {C\over \h} E^{1\over 1+\h}
\ee
where $C$ is a suitable constant and $\e$ is an extra ultraviolet cut-off
; that is the density of states vanishes with a critical exponent.

 \section{Conclusions}

We have analyzed for the first time
chiral Dirac fermions in presence of a momentum cut-off
and short range disorder,  extending previous results in which only delta correlated disorder
without ultraviolet cut-off was considered.
The model provides a more realistic description 
in view
of applications to condensed matter models, and  is free from any ultraviolet
divergence. We have shown that the density of states is anomalous with a critical exponent function
of the disorder and that 
the conductivity is exactly universal {\it only} when the ultraviolet cut-off is removed;
this may have implications for the physics of graphene in which a natural ultraviolet cut-off is provided by the honeycomb lattice.


\end{document}